\documentclass[11pt]{article}
\usepackage{amsxtra,amssymb,amsthm,amsmath,latexsym}
\usepackage{graphicx}

\newtheorem{theorem}{Theorem}

\begin{document}
\date{}
\title{Many-body wave scattering by small bodies\newline and applications 
II.}

\author{A. G. Ramm\footnotemark[1]\quad and \quad A. 
Rona\footnotemark[3]\\
Mathematics Department, Kansas State University,\\ Manhattan,
KS66506, USA\\ 
Department of Engineering, University of Leicester,\\
Leicester LE1 7RH, UK}

\renewcommand{\thefootnote}{\fnsymbol{footnote}}
\footnotetext[1]{Email:  ar45@leicester.ac.uk}
\footnotetext[3]{Corresponding author. Email: ramm@math.ksu.edu}

\maketitle

\begin{abstract} 
The many-body wave scattering problem is studied in the
case where the bodies are small, $ka \ll 1$, where $a$ is the
characteristic size of a body. The limiting case when $a \to 0$ and the
total number of the small bodies is $M = O ( a^{2-\kappa} )$ is
studied. 

\textbf{keywords: } wave scattering, small scatterers, many-body scattering problem

\end{abstract}

\section{Introduction} The many-body scattering problem in the case of
small scatterers embedded in an inhomogeneous medium has been solved in
\cite{r509} and \cite{r536} under the following assumptions:
\begin{equation}
\label{eq:assumptions}
ka \ll 1, \quad d = O \big{(} a^{\frac{2-k}{3}} \big{)}, 
\quad \zeta_{m} = \frac{h(\mathbf{x}_{m}) }{a^{\kappa}}, 
\end{equation}
where $a$ is the characteristic size of the small bodies,
$\kappa = 2\pi/\lambda = \omega/c_{0}$ is the wave number and $c_{0}$ is
the wave speed in free space, $\kappa \in ( 0, 1 )$ is a
parameter one can choose as one wishes, $d$ is the distance between
neighboring particles, $h( \mathbf{x} )$ is a
piecewise-continuous function in a bounded domain 
$D \subset \mathbb{R}^{3}$ with a
smooth boundary $S$, $\Im h = h_{2} \leq 0$, $h = h_{1} + i h_{2}$,
$\mathbf{x}_{m} \in D_{m}$ is an arbitrary point, $D_{m}$ is a small body,
$S_{m}$ is its surface, $\mathbf{N}$ is the unit normal to $S_{m}$, $1
\leq m \leq M$, $M$ is the total number of the embedded small bodies in
$D$, the unit normal $\mathbf{N}$ points out of $D_{m}$, $\zeta_{m}$ is 
the boundary impedance in the boundary condition:  
\begin{equation} 
\label{eq:array}
\frac {\partial u}{\partial \mathbf{N}} = \zeta_{m} u\qquad \text{on}\quad 
S_{m}, 
\quad 1
\leq m \leq M; \qquad u = u_{M},  
\end{equation} and the
distribution of small bodies in $D$ is defined as  
\begin{equation}
\label{eq:smallbodies}
\mathcal{N} ( \Delta ) := \Sigma_{D_{m} \subset \Delta} 1 =
\frac{1}{a^{2-\kappa}} \int_{\Delta} N(\mathbf{x}) d
\mathbf{x}[ 1 + o ( 1 ) ], \quad a \to 0,
\end{equation} where $\mathcal{N}( \Delta
)$ is the number of small bodies in an arbitrary subdomain $\Delta
\subset D$, $N(\mathbf{x}) \geq 0$ is a piecewise-continuous
function, and for simplicity it is assumed that $D_{m} = B (
\mathbf{x}_{m}, a )$ is a ball centred at the point
$\mathbf{x}_{m}$, of radius $a$. The scattering problem, solved in
\cite{r509}-\cite{r536}, consisted of finding the solution to the equation 
\begin{equation} 
\label{eq:goveqn} 
[ \nabla^{2} + k^{2} n_{0}^{2} ( \mathbf{x}
) ] u_{m} = 0 \quad \text{in}\quad \mathbb{R}^3\setminus 
\bigcup_{m=1}^{M}D_m, \quad
\end{equation} 
satisfying boundary conditions
\eqref{eq:array} and the radiation condition  
\begin{equation} 
\label{eq:radcondition} 
u_{M} =
u_{0} + v_{M}, \quad \frac {\partial v_{M}}{\partial r} - ik v_{M} 
= o (\frac{1}{r} ), \quad r:= | \mathbf{x} | \to \infty.
\end{equation}

Here, $u_{0}$ is the solution to problem 
\eqref{eq:goveqn}-\eqref{eq:radcondition}  in the absence of the 
embedded particles, i.e.,
the solution for the problem with $M=0$:  
\begin{equation}
\label{eq:helmholtz} 
[\nabla^{2} + k^{2} n_{0}^{2} ( \mathbf{x} ) ] u_{0} = 0 \quad
\quad \text{in}\quad \mathbb{R}^{3},  
\end{equation}
where $n_0^2(x)$ is the refraction coefficient in the absence of
embedded particles, $n_0^2(x)=1$ in the region 
$D':=\mathbb{R}^{3}\setminus D$, and
\begin{equation}
\label{eq:radcondition2}
u_{0} = e^{ik \alpha \cdot \mathbf{x}} + v_{0}, 
\quad \frac{\partial v_{0}}{\partial r} - i k v_{0} = 
o \bigg{(}\frac{1}{r}\bigg{)}, 
\quad r \to \infty,
\end{equation}
where $\alpha \in S^{2}$ is the direction of propagation of
the  incident plane wave, and $S^{2}$ is 
the unit sphere in $R^{3}$. 

It was proved in \cite{r509},\cite{r536}, that,  as $a \to 0$,
the limiting field $u$ does exist and solves the equation
\begin{equation}
\label{eq:helmholtzfreefield}
[ \nabla^{2} + k^{2} n^{2} ( \mathbf{x} ) ] u = 0 \quad \text{in}\quad 
\mathbb{R}^{3}, \quad
\end{equation}
where
\begin{equation}
\label{eq:refractioncoeff}
n^{2} ( \mathbf{x} )\equiv  n_{0}^{2} ( \mathbf{x} ) - 4 \pi k^{-2} h 
(\mathbf{x}) N ( \mathbf{x} ).
\end{equation}

Therefore, in the limit $a \to 0$, under the constraints
\eqref{eq:assumptions} - \eqref{eq:smallbodies}, the limiting medium,
obtained by the embedding of many small particles, has the refraction
coefficient $n^{2} ( \mathbf{x} )$, given by \eqref{eq:refractioncoeff}.  
Since the functions $h ( \mathbf{x} )$ and $N ( \mathbf{x} )$ are at our
disposal, subject to the restrictions $N ( \mathbf{x} ) \geq 0$, $\Im h (
\mathbf{x} ) \leq 0$, it is possible to create any desired refraction
coefficient $n^{2} ( \mathbf{x} )$, $\Im n^{2} ( \mathbf{x} ) \geq 0$, by
choosing $h ( \mathbf{x} )$ and $N ( \mathbf{x} )$ suitably.

 It is assumed that the term "piecewise-continuous" function $f$ in
this paper means that the set $\mathcal{M}$ of discontinuities of $f$ is
of Lebesgue's measure zero and, if $\mathcal{S}$ is a subset of this set
such that $f$ is unbounded on $\mathcal{S}$, $f|_{\mathcal{S}} = \infty$, 
then $f$ grows not too fast as $\mathbf{x}$ tends to $\mathcal{S}$:
 \begin{equation}
\label{eq:lebesegne} |f (\mathbf{x}) | \leq \frac{c}{[ \text{dist} (
\mathbf{x}, \mathcal{S} ) ]^{\nu}}, \quad 0 \le \nu < 3, \quad
c=\mbox{const} \ge 0, \end{equation} 
so that the integral  $\int_{D} f ( \mathbf{x} ) d \mathbf{x}$ exists as 
an improper integral.

This paper is closely related to \cite{r509}, and its goal  
is to develop a theory, similar to the one in \cite{r509}, for a 
different governing equation:
\begin{equation}
L_{0}u_{0} := \nabla \cdot ( c^{2} ( \mathbf{x} ) \nabla u_{0} ) + 
\omega^{2} u_{0} = 0 \quad \mbox{in} \quad R^{3}, \quad
\label{eq:wavevariablespeed}
\end{equation}
where the wave speed $c ( \mathbf{x} ) = c_{0}=const$ in $D^{\prime} := 
R^{3}\backslash D$, the complement of $D$ in $R^{3}$, and  $c ( \mathbf{x} 
)$ 
is a smooth and strictly positive function  in $D$. The speed
$c ( \mathbf{x})$, in general, has $S$ as 
its discontinuity surface. In this case equation 
\eqref{eq:wavevariablespeed} is 
understood in the distributional sense as an integral identity:
\begin{equation}
\int_{R^{3}} ( - c^{2} ( \mathbf{x} ) \nabla \phi \nabla u_{0} + 
\omega^{2} \phi u_{0} ) d \mathbf{x} = 0 \quad \forall \phi 
\in C_{0}^{\infty} ( R^{3} ). \quad
\label{eq:integralidentity}
\end{equation}

Alternatively, one may understand equation \eqref{eq:wavevariablespeed} 
as the following transmission problem:
\begin{align}
\label{eq:infield}
L_{0}u_{0}^{+} &= 0 \quad \text{in}\quad  D, \quad u_{0}^{+} = u_{0} 
\quad \text{in}\quad  D, \\
\label{eq:outfield}
L_{0}u_{0}^{-} &= 0 \quad \text{in}\quad D', \quad u_{0}^{-} = u_{0} 
\quad \text{in} \quad D' 
\end{align}
\begin{equation}
u_{0}^{+} = u_{0}^{-}, \quad c^{2}_{+} ( \mathbf{x} ) \frac {\partial 
u{0}}{\partial \bf{N}^{+}}  
= c^{2}_{-} ( \mathbf{x} ) \frac {\partial u_{0}}{\partial \bf{N}^{-}} 
\quad \mbox{on}\quad S.
\label{eq:transmissioncondition}
\end{equation}

The transmission conditions \eqref{eq:transmissioncondition} together with
equations \eqref{eq:infield} and \eqref{eq:outfield} are equivalent to
problem \eqref{eq:integralidentity} . Existence and uniqueness of the
solution to \eqref{eq:infield} - \eqref{eq:transmissioncondition} was
proved in \cite{r163}.

The scattering problem we are interested in  can be stated as follows:
\begin{align}
\label{eq:wave-equation}
L_{0}u &= 0\quad \text{in} \quad R^{3}\backslash \bigcup_{m=1}^{M} D_{m}; 
\quad u=u_{M}, \\
\label{eq:impedancebc}
\frac {\partial u}{\partial \bf{N}} &= \zeta_{m} u \quad \text{on}\quad 
S_{m}, \quad 1 \leq m \leq M, 
\quad \\
\label{eq:sommerfieldbc}
u &= u_{0} + v, \quad v_{r} - i k v = o ( \frac{1}{r} ), \quad r \to 
\infty. 
\end{align}

In section \ref{sec:scattering} problem 
\eqref{eq:wave-equation}  - \eqref{eq:sommerfieldbc}  
is investigated and the limiting behavior of $u$ as $a \to 0$
is found.

 We conclude this Introduction by a brief derivation of the governing 
equation \eqref{eq:wavevariablespeed}. 

The starting point is the Euler equation:
\begin{equation} \dot{\mathbf{v}} + (\mathbf{v}, \nabla ) \mathbf{v} =
- \frac{\nabla p}{\rho}, 
\label{eq:euler} \end{equation} 
where $\mathbf{v}$
is the velocity vector of the sound wave, $p=p ( \rho )$ is the static
pressure, $\rho$ is the density, and
\begin{equation} 
\label{eq:soundspeed} 
\nabla p = c^{2}
( \mathbf{x} ) \nabla \rho, 
\end{equation}
where $c( \mathbf{x} )$ is the sound speed.

Let the material in $D$ be initially at rest,  $\mathbf{v} = \mathbf{v}
(\mathbf{x},t)$ be a small perturbation of the equilibrium zero 
velocity, the density be of the form
$\rho = \rho_{0} + \psi (\mathbf{x},t)$, where $\rho_{0}$ is the
equilibrium density of the material, which is assumed to be constant, and 
$\psi$ and
$\mathbf{v}$ are small quantities of the same order of smallness.

The continuity equation is
\begin{equation}
\dot{\psi} = - \nabla \cdot ( \rho_{0} \mathbf{v} ),
\label{eq:continuity}
\end{equation}
where the term $\nabla \cdot( \psi \mathbf{v} )$ of the higher order of 
smallness is neglected. Differentiating \eqref{eq:continuity}  
with respect to time yields
\begin{equation}
\label{eq:dcontinuitydt}
\ddot{\psi} = - \nabla \cdot ( \rho_{0} \dot{\mathbf{v}} ).
\end{equation}

Under the same assumptions about 
$\rho = \rho_{0} + \psi ( \mathbf{x}, t )$ 
and $\mathbf{v}$, the term $( \mathbf{v}, \nabla ) 
\mathbf{v}$ in \eqref{eq:euler}  is of the higher order of smallness and 
is therefore neglected. Multiplying \eqref{eq:euler}  by $\rho$ and 
neglecting 
the term $\psi \dot{\mathbf{v}}$ of higher order of smallness 
yields the acoustic momentum equation
\begin{equation}
\label{eq:acousmomentum}
\rho_{0} \dot{\mathbf{v}} = - \nabla p.
\end{equation}

Substituting \eqref{eq:soundspeed}  in \eqref{eq:acousmomentum}  gives
\begin{equation}
\label{eq:acousmomentum2}
\rho_{0} \dot{\mathbf{v}} = - c^{2} ( \mathbf{x} ) \nabla \psi,
\end{equation}
where the relation $\nabla \rho = \nabla \psi$ was used. This relation 
is exact for a constant $\rho_{0}$. 

Substituting 
\eqref{eq:acousmomentum2}  in \eqref{eq:dcontinuitydt}  yields
\begin{equation}
\label{eq:wave-equation-variable-phase-speed}
\ddot{\psi} - \nabla \cdot ( c^{2} ( \mathbf{x} ) \nabla \psi ) = 0.
\end{equation}
If $\psi = e^{-i \omega t} u$, 
then \eqref{eq:wave-equation-variable-phase-speed}  reduces to 
equation \eqref{eq:wavevariablespeed} .

\section{The scattering problem}
\label{sec:scattering}
In this Section, problem \eqref{eq:wave-equation}  - 
\eqref{eq:sommerfieldbc}  is studied. Assumptions \eqref{eq:assumptions} 
and  \eqref{eq:smallbodies}  are still valid. 

Let $G$ be the Green's 
function 
for the operator $L_{0}$:
\begin{equation}
\label{eq:greenfunction}
L_{0} G ( \mathbf{x}, \mathbf{y} ) = - \delta ( \mathbf{x} - \mathbf{y} )
\quad \text{in}\quad \mathbb{R}^{3},
\end{equation}

$G$ satisfies the radiation condition
\begin{equation}
\label{eq:Gradcondition}
\frac{\partial G}{\partial | \mathbf{x} |} - ikG = o ( \frac{1}{| 
\mathbf{x} |} )\quad  \mbox{as} \quad | \mathbf{x} | \to \infty.
\end{equation}

The following result from \cite{r371} will be used.

\begin{theorem}
In a neighborhood of a point of smoothness of $c ( \mathbf{x} )$ one has
\begin{equation}
G ( \mathbf{x}, \mathbf{y} ) = \frac{1}{4\pi | \mathbf{x} - \mathbf{y} | 
c ( \mathbf{x} )} ( 1 + o (1) ), \quad | \mathbf{x} - \mathbf{y} | \to 0. 
\quad
\label{eq:Gfreefield}
\end{equation}

In a neighborhood of the point $x \in S$, where $S$ us a smooth 
discontinuity surface of $c ( \mathbf{x} )$, one has
\begin{equation}
\label{eq:GinDcomplement}
G ( \mathbf{x}, \mathbf{y} ) = \left \{
\begin{matrix}
\frac{1}{4\pi c_{+} ( \mathbf{x} )} [ r_{\mathbf{x}\mathbf{y}}^{-1} + b 
R^{-1} + o (1)], \quad \mathbf{y} \in\quad  D,\\
\frac{1}{4\pi c_{-} ( \mathbf{x} )} [ r_{\mathbf{x}\mathbf{y}}^{-1} - 
b R^{-1} + o (1)], \quad \mathbf{y} \in\quad D'.
\end{matrix}
\right .
\end{equation}
where
\begin{align}
\label{eq:29} 
b:=& \frac{c_{+} ( \mathbf{x} ) - c_{-} ( \mathbf{x} )}{c_{+} ( \mathbf{x} ) - c_{-} ( \mathbf{x} )}, \quad r_{\mathbf{x}\mathbf{y}} := | \mathbf{x} - \mathbf{y} |, \quad R = \sqrt{\rho^{2} + ( | x_{3} | + | y_{3} | )^{2}},
\\
\label{eq:30}
\rho =& \sqrt{( x_{1} - y_{1} )^{2} + ( x_{2} - y_{2} )^{2}}. \quad
\end{align}

The origin of the local coordinate system lies on $S$, the plane $x_{3} =
0$ is tangent to $S$, $c_{+} ( \mathbf{x} )$ and $c_{-} ( \mathbf{x} )$
are the limiting values of $c ( \mathbf{x} )$ when $\mathbf{x} \to S$ from
inside (and outside) of $D$.
\end{theorem}

In \cite{r371}, the operator $L_{0}$ corresponds to the case $\omega = 0$.
However, in \cite{r509} it is proved that adding to $L_{0}$ a term $q (
\mathbf{x} ) G ( \mathbf{x}, \mathbf{y} )$ with a bounded function $q$
does not change the main term of the asymptotic of $G$ as $\mathbf{x} \to
\mathbf{y}$.

The solution to problem \eqref{eq:wave-equation} -
\eqref{eq:transmissioncondition} is sought in the form 
\begin{equation} u = u_{0} + \sum_{m=1}^{M} \int_{S_{m}} G ( \mathbf{x}, t
) \sigma_{m} ( t )  dt. 
\label{eq:31} 
\end{equation}

For any $\sigma_{m} \in L^{2} ( S_{m} )$, the function $u$,
defined in  \eqref{eq:31}, 
solves
equation \eqref{eq:wave-equation} and satisfies the radiation condition
\eqref{eq:transmissioncondition} , since $G$ does.
Therefore, \eqref{eq:31} will be solution to
problem \eqref{eq:wave-equation} - \eqref{eq:transmissioncondition} if
$\sigma_{m}$ are such that the boundary conditions \eqref{eq:impedancebc}
are satisfied. Uniqueness of the solution to problem
\eqref{eq:wave-equation} - \eqref{eq:transmissioncondition} follows from
essentially the same arguments as in \cite{r509}, see the proof of Theorem 
1 in \cite{r509}. 

The boundary conditions \eqref{eq:transmissioncondition}
imply 
\begin{equation} u_{eN} + \frac{A_{m}\sigma_{m} - \sigma_{m}
c_{m}^{-1}}{2} = \zeta_{m} u_{e} + \zeta_{m} \int_{S_{m}} G ( s, t )
\sigma_{m} ( t ) dt, 
\label{eq:32} 
\end{equation} 
where 
$$c_{m} := c (\mathbf{x}_{m} ), \qquad \zeta_{m} = 
h ( \mathbf{x}_{m} ) / a^{\kappa},$$ 
and
\begin{equation} u_{e} ( \mathbf{x} ) := u_{e}^{( m )} := u_{0} (
\mathbf{x} ) + \sum_{m' \neq m} \int_{S_{m}'} G ( \mathbf{x}, t )
\sigma_{m'} ( t ) dt. 
\label{eq:33} 
\end{equation} 
The field $u_{e}^{( m)}$ is called the effective (self-consistent) field. 
It is the field acting on the
$m$-th particle from all other particles and from the incident field
$u_{0}$. 

The operator $A_{m}$ is the operator of the normal derivative of 
the single-layer
potential 
$$T\sigma_{m} := \int_{S_{m}} G ( \mathbf{x}, t )
\sigma_{m} ( t ) dt$$ 
at the boundary $S$,  and 
\begin{equation}
\frac{\partial T\sigma_{m}}{\partial \bf{N}^{-}} = \frac{A \sigma_{m} -
\sigma_{m} c^{-1} ( \mathbf{x}_{m} )}{2}, \quad A \sigma_{m} =
\int_{S_{m}} \frac{\partial G ( \mathbf{s}, t ) }{\partial \bf{N}_{s}}
\sigma_{m} ( t ) dt, \quad \mathbf{s} \in S_{m}, \label{eq:34}
\end{equation}

In equation \eqref{eq:34}, $\frac{\partial T\sigma_{m}}{\partial 
\bf{N}^{-}}$ 
is the limiting value of the normal derivative on $S_{m}$ from outside 
of $D_{m}$. 
Equation \eqref{eq:34} is well known from the potential theory in the 
case $c ( \mathbf{x})  = 1$ and 
$G ( \mathbf{x}, \mathbf{y} ) = \frac{e^{i\omega  r_{\mathbf{x} \mathbf{y}} }}
{4\pi r_{\mathbf{x}\mathbf{y}}} $, $ 
r_{\mathbf{x}\mathbf{y}}:=|\mathbf{x}-\mathbf{y}|$. 

If $c ( \mathbf{x} ) \neq 1$, then by Theorem 1 one 
may 
consider $T$ and $A$ as the operators, corresponding to $c ( \mathbf{x} ) 
= 1$, 
divided by $c ( \mathbf{x}_{m} )$, because $c ( \mathbf{s} )$
is assumed smooth in $D$, and, therefore, it  
varies negligibly on the small distances of the order $a$.

The basic idea of solving the many-body scattering problem
\eqref{eq:wave-equation} - \eqref{eq:sommerfieldbc} is similar to the idea
originally used in \cite{r509}. 

The approach is to reduce the solution of the many-body scattering problem 
by small bodies to finding some numbers, rather than the unknown functions 
$\sigma_m$, $1\leq m \leq M$. If $M$ is very large, it is practically 
impossible to use the usual system of boundary integral equations for 
finding the unknown $\sigma_m$.

let us rewrite equation \eqref{eq:31} as follows:

\begin{equation}
\label{eq:35}
u = u_{0} ( \mathbf{x} ) + \sum_{m=1}^{M} G ( \mathbf{x}, \mathbf{x}_{m} ) 
Q_{m} + \sum_{m=1}^{M} \int_{S_{m}} [ G ( \mathbf{x}, t ) 
- G ( \mathbf{x}, \mathbf{x}_{m} )] \sigma_{m} ( t ) dt,
\end{equation}
where
\begin{equation}
\label{eq:36}
Q_{m} := \int_{S_{m}} \sigma_{m} ( t ) dt,
\end{equation}
and prove,  as it is done in \cite{r509}, that
\begin{equation}
\label{eq:37}
\bigg{|} G ( \mathbf{x}, \mathbf{x}_{m} ) Q_{m}\bigg{|} \gg 
\bigg{|}\int_{S_{m}} [ G ( \mathbf{x}, t ) - 
G ( \mathbf{x}, \mathbf{x}_{m} ) ] \sigma_{m} ( t ) dt \bigg{|}, 
\qquad a \to 0; \qquad | \mathbf{x} - \mathbf{x}_{m} | > a.
\end{equation}
Thus, the solution $u$ can be written as
\begin{equation}
\label{eq:38}
u = u_{0} ( \mathbf{x} ) + \sum_{m=1}^{M} G ( \mathbf{x}, \mathbf{x}_{m} ) 
Q_{m}, \quad | \mathbf{x} - \mathbf{x}_{m} | > a, \qquad
\end{equation}
with the error that tends to zero as $a \to 0$. 

Consequently, the scattering problem is solved if the numbers $Q_{m}$, $1
\leq m \leq M$, are found. This simplifies the solution of the 
many-body scattering problem drastically,
because equation \eqref{eq:31} requires the knowledge of 
the functions $\sigma_{m} ( t )$, $1 \leq m \leq M$, rather than the 
numbers $Q_{m}$, in order to find the solution $u$ of the scattering 
problem.

The next step is to derive the main term of the asymptotics of $Q_{m}$ as
$a \to 0$.

To do this, we integrate \eqref{eq:32}  over $S_{m}$ and neglect the 
terms of 
the higher order of smallness as $a \to 0$. 

One has:
\begin{equation}
\int_{S_{m}} u_{eN} d\mathbf{s} = \int_{D_{m}} \nabla^{2} u_{e} 
d\mathbf{x} = ( \nabla^{2} u_{e} ) ( \mathbf{x}_{m} ) | D_{m} |, \quad | 
D_{m} | = \frac{4\pi a^{3}}{3}, \quad
\label{eq:39}
\end{equation}
where the Gauss divergence theorem was applied 
and a mean value formula for  the integral over $D_m$ was used.

Furthemore,
\begin{equation}
\int_{S_{m}} A \sigma_{m} d \mathbf{s} = - \frac{1}{c_{m}} \int_{S_{m}} 
\sigma_{m} d\mathbf{s} = - \frac{Q_{m}}{c_{m}},
\label{eq:40}
\end{equation}
where (cf  \cite{r509})
\begin{equation}
\int_{S_{m}} A \sigma d\mathbf{s} :=\frac{1}{c_{m}} \int_{S_{m}} 
d\mathbf{s}\int_{S_{m}} 
\frac{\partial}{\partial N_{s}} \frac{1}{4\pi r_{\mathbf{s}t}} 
\sigma ( t ) dt = -\frac{1}{c_{m}} \int_{S_{m}} \sigma ( t ) dt.
\label{eq:41}
\end{equation}

Thus, integrating \eqref{eq:32}  over $S_{m}$ yields
\begin{equation}
\nabla^{2} u_{e} ( \mathbf{x}_{m} ) | D_{m} | - c^{-1}_{m} Q_{m} = 
\zeta_{m} u_{e} ( x_{m} ) | S_{m} | + \frac{\zeta_{m}}{c_{m}} 
\int_{S_{m}} dt \sigma_{m} ( t ) \int_{S_{m}} d\mathbf{s} 
\frac{1}{4\pi r_{\mathbf{s}t}},
\label{eq:42}
\end{equation}
where $|S_{m}| = 4\pi a^{2}$ is the surface area of the sphere $S_m$ and 
formula \eqref{eq:Gfreefield}  was used, namely,  
we have replaced $G ( \mathbf{s}, t )$ by $\frac{1}{4\pi r_{\mathbf{s}t}} 
\frac{1}{c ( \mathbf{s} )}$ using the smallness of $D_{m}$, and we have 
replaced $c ( \mathbf{s} )$ by $c ( \mathbf{x}_{m} ) = c_{m}$ because
$|x_m-s|\leq a$, and $a$ is small. 

Using the identity
\begin{equation}
\int_{S_{m}} \frac{d\mathbf{s}}{4\pi r_{\mathbf{s}t}} = a \quad 
\mbox{if}\quad 
| \mathbf{s} - \mathbf{x}_{m} | = a \quad \mbox{and}\quad | t - 
\mbox{x}_{m} | = a,
\label{eq:43}
\end{equation}
one gets from \eqref{eq:42}  the following relation: 
\begin{equation}
Q_{m} ( c_{m}^{-1} + c_{m}^{-1} \zeta_{m} a ) = - 4 \pi \zeta_{m} u_{e} 
( \mathbf{x}_{m} ) a^{2} + O ( a^{3} ).
\label{eq:44}
\end{equation}

If $a \to 0$ and $\kappa\in (0,1)$, then 
$$\zeta_{m} a = h ( \mathbf{x}_{m} ) a^{1-\kappa} = o ( 1 ),\qquad a\to 
0,$$ 
the term $O ( a^{3} )$ in \eqref{eq:44} can be neglected, and one gets the 
main 
term of the asymptotics of $Q_{m}$ as $a \to 0$, namely:
\begin{equation}
Q_{m} = - 4\pi h ( \mathbf{x}_{m} ) u_{e} ( \mathbf{x}_{m} ) c ( 
\mathbf{x}_{m} ) a^{2-\kappa} [1 + o (1)], \qquad a \to 0. 
\label{eq:45}
\end{equation}

Therefore \eqref{eq:33} , \eqref{eq:38}  and \eqref{eq:45}  yield
\begin{equation}
u_{e} ( \mathbf{x} ) = u_{0} ( \mathbf{x} ) - 4\pi \sum_{m^{\prime}\neq m} 
G ( \mathbf{x}, \mathbf{x}_{m} ) h ( \mathbf{x}_{m^{\prime}} ) 
u_{e} ( \mathbf{x}_{m^{\prime}} ) c ( \mathbf{x}_{m^{\prime}} ) 
a^{2-\kappa} [ 1 + o ( 1 ) ].
\label{eq:46}
\end{equation}

Taking $\mathbf{x} = \mathbf{x}_{m}$ and neglecting $o(1)$ term in 
\eqref{eq:46},  
one gets a linear algebraic system for the unknown quantities 
$u_{m} := u_{e} ( \mathbf{x}_{m} )$, $1\leq m\leq M$, 
\begin{equation}
u_{m} = u_{0m} - 4\pi \sum_{m^{\prime} \neq m} G ( \mathbf{x}_{m}, 
\mathbf{x}_{m^{\prime}} ) h ( \mathbf{x}_{m^{\prime}} ) 
c ( \mathbf{x}_{m^{\prime}} ) u_{m^{\prime}} a^{2-\kappa}.
\label{eq:47}
\end{equation}

Let us now derive and use a generalization of the result proved originally 
in \cite{r509}. This generalization is formulated as Theorem 2 below.

Consider the sum 
\begin{equation} I = \lim_{a \to 0}
a^{2-\kappa} \sum_{m=1}^{M} f ( \mathbf{x}_{m} ), \label{eq:48}
\end{equation} 
where the points $\mathbf{x}_{m}$ are distributed in $D$
according to \eqref{eq:smallbodies}.

Assume that $f ( \mathbf{x} )$ is piecewise-continuous
and \eqref{eq:lebesegne} holds. If $f$ is unbounded, that is, the set
$\mathcal{S}$ is not empty, then the sum \eqref{eq:48} is understood 
as follows: 
 \begin{equation} I := \lim_{\delta \to 0} \lim_{a \to 0} a^{2-\kappa}
\sum_{m=1, dist ( \mathbf{x}_{m}, \mathcal{S} ) \geq \delta}^{M} f (
\mathbf{x}_{m} ). 
\label{eq:49} \end{equation}

\begin{theorem} 
Under the above assumptions, there exists the limit \eqref{eq:48}  and
\begin{equation}
\lim_{a \to 0} a^{2-\kappa} \sum_{m=1}^{M} f ( \mathbf{x}_{m} ) = 
\int_{D} f ( \mathbf{x} ) N ( \mathbf{x} ) d \mathbf{x}.
\label{eq:50}
\end{equation}
\end{theorem}
Proof of Theorem 2 is given at the end of this paper.

Applying Theorem 2 to the sum \eqref{eq:46}  one obtains the following 
result:

\begin{theorem} 
There exists the limit:
$$\lim_{a \to 0} u_{e} ( \mathbf{x} ) := u ( 
\mathbf{x} ),$$ 
and the limiting function solves the equation:
\begin{equation}
u ( \mathbf{x} ) = u_{0} ( \mathbf{x} ) - 4\pi \int_{D} 
G ( \mathbf{x}, \mathbf{y} ) h ( \mathbf{y} ) 
c ( \mathbf{y} ) N ( \mathbf{y} ) u ( \mathbf{y} ) d \mathbf{y}.
\label{eq:51}
\end{equation}
\end{theorem}
Applying operator $L_{0}$, defined in \eqref{eq:wavevariablespeed} , 
to \eqref{eq:51}  and using the relations
\begin{equation} 
L_{0} G = - \delta ( \mathbf{x} - \mathbf{y} ), \quad L_{0} u_{0} = 0, \quad
\label{eq:51b}
\end{equation}
one obtains the following new equation for the limiting effective field 
$u$: 
\begin{equation}
L_{0} u = 4\pi h ( \mathbf{x} ) c ( \mathbf{x} ) N ( \mathbf{x} ) u.
\label{eq:52}
\end{equation}
This equation can be written as:
\begin{equation}
Lu:=\nabla \cdot ( c^{2} ( \mathbf{x} ) \nabla u ) + \omega^{2} u - 
4\pi h ( \mathbf{x} ) c ( \mathbf{x} ) N ( \mathbf{x} ) u = 0.
\label{eq:52b}
\end{equation}

Therefore, embedding many small particles into $D$ and assuming
\eqref{eq:assumptions} - \eqref{eq:smallbodies} , one obtains in the limit
$a \to 0$ a medium with essentially different properties described by
the new equation \eqref{eq:52b}.
$$ $$
Let us now prove Theorem 2.

{\bf Proof of Theorem 2.} 

Let $\mathcal{S}$ be the subset of the set of discontinuities of $f$
on which $f$ is unbounded, let the assumption \eqref{eq:lebesegne} hold, 
and let
\begin{equation}
\label{eq:53}
D_{\delta} := \{ \mathbf{x} : \mathbf{x} \in D, 
dist ( \mathbf{x}, \mathcal{S} ) \geq \delta \}.
\end{equation}

Consider a partition of $D_{\delta}$ into a union of small cubes
$\Delta_{p}$, centered at the points $\mathbf{y}_{p}$, with the side $b =
a^{1/3}$. One has 
\begin{equation} \label{eq:54} \begin{split}
a^{2-\kappa} \sum_{m=1, dist ( \mathbf{x}_{m}, \mathcal{M} ) \geq
\delta}^{M} f ( \mathbf{x}_{m} ) &= \sum_{p} f ( \mathbf{y}_{p} ) [ 1 + o
( 1 ) ] a^{2-\kappa} \sum_{\mathbf{x}_{m} \in \Delta_{p}} 1 \\ &= \sum_{p}
f ( \mathbf{y}_{p} ) N ( \mathbf{y}_{p} ) | \Delta_{p} | [ 1 + o ( 1 ) ]
\\ & \rightarrow \int_{D_{\delta}} f ( \mathbf{y} ) N ( \mathbf{y}
) d\mathbf{y}\quad \text{as} \quad a \to 0. \end{split} 
\end{equation}

Here in the second sum we replaced $f ( \mathbf{x}_{m} )$ by 
$f ( \mathbf{y}_{p} )$ for all points $\mathbf{x}_{m} \in \Delta_{p}$. 
This is done with the error $o ( 1 )$ as $a \to 0$, because $f$ is 
continuous in 
$D_{\delta}$. In the third sum we have used formula \eqref{eq:smallbodies}  
for $\Delta = \Delta_{p}$. The last conclusion, namely, 
the existence of the limit as $a\to 0$, follows from the 
known result: the Riemannian sum of a piecewise-continuous bounded in 
$D_{\delta}$ function $f ( \mathbf{x} ) N ( \mathbf{x} )$ 
converges to the integral 
$\int_{D_\delta} f ( \mathbf{x} ) N ( \mathbf{x} ) d\mathbf{x}$ 
if $\max_{p} diam \Delta_{p} \to 0$. In our case
\begin{equation}
\label{eq:54b}
diam \Delta_{p} = \sqrt{3} a^{1/3} \to 0 \quad \mbox{as} \quad a \to 0,
\end{equation}
so formula \eqref{eq:54}  follows.

From the assumption \eqref{eq:lebesegne}  with $\nu < 3$ one concludes 
that
\begin{equation}
\label{eq:55}
\lim_{\delta \to 0} \int_{D_{\delta}} f ( \mathbf{x} ) N ( \mathbf{x} ) 
d\mathbf{x} = \int_{D} f ( \mathbf{x} ) N ( \mathbf{x} ) d\mathbf{x}.
\end{equation}

The integral on the right in \eqref{eq:55} exists as an improper integral
if $\nu$ is less than the dimension of the space, i.e., $\nu < 3$.
Therefore, formula \eqref{eq:50} is established. 

Theorem 2 is proved.     \hfill $\Box$

\end{document}